\documentclass{aa}
\def\e{et~al.\ }

\def\r{R_{\rm m}}
\newcommand{\be}{\begin{equation}}
\newcommand{\ee}{\end{equation}}
\newcommand{\bdm}{\begin{displaymath}}
\newcommand{\edm}{\end{displaymath}}

\begin{document}

   \title{On the duration of the subsonic propeller state of neutron
stars in wind-fed mass-exchange close binary systems}

   \author{Nazar R. Ikhsanov\inst{1,2}}


   \institute{Max-Planck-Institut f\"ur Radioastronomie, Auf dem
              H\"ugel 69, D-53121 Bonn, Germany\\
              \email{ikhsanov@mpifr-bonn.mpg.de}
              \and
              Central Astronomical Observatory of
              the Russian Academy of Sciences at Pulkovo,
              Pulkovo 65--1, 196140 Saint-Petersburg, Russia}

   \date{Received ; accepted }

\titlerunning{On the duration of the subsonic propeller state}

    \abstract{
The condition for the {\it subsonic propeller}\,$\rightarrow$\,{\it
accretor} state transition of neutron stars in wind-fed mass-exchange
binary systems is discussed. I show that the value of the break
period, at which the neutron star change its state to {\it accretor},
presented by Davies \& Pringle (\cite{dp81}) is underestimated by a
factor of 7.5. The correct value is $P_{\rm br} \simeq\ 450\
\mu_{30}^{16/21}\ \dot{M}_{15}^{-5/7}\ (M/M_{\sun})^{-4/21}$\,s. This
result forced us to reconsider some basic conclusions on the
efficiency of the propeller spindown mechanism.
   }

   \maketitle


   \section{Introduction}

The sequence of states which a magnetized neutron star in a wind-fed
mass-exchange binary system follows as it spins down from the
initially very short periods can be expressed in the form of the
following chain: {\it ejector} $\rightarrow$ {\it propeller}
$\rightarrow$ {\it accretor}. This classification, first
suggested by Shvartsman (\cite{sh70}), reflects three different
evolutionary stages and three different mechanisms of energy release
responsible for the neutron star emission.

The spindown of a neutron star in the state of ejector is governed
by the canonical spin-powered pulsar mechanism. The spindown power
dominates the star energy budget and is spent predominantly to the
generation of magneto-dipole waves and particle acceleration. The
pulsar-like spindown ceases when the pressure of the material
being ejected by the neutron star can no longer balance the pressure
of the surrounding gas. The stellar wind plasma penetrates into the
neutron star light cylinder and interacts with the star
magnetosphere. This corresponds to the neutron star state
transition: {\it ejector} $\rightarrow$ {\it propeller}.

Neutron star in the state of propeller is spinning down due to the
interaction between its fast rotating magnetosphere and the
surrounding material. Davies \e (\cite{dfp79}) and Davies \&
Pringle (\cite{dp81}) have shown that during this
state the star magnetosphere is surrounded by a spherical quasi-static
envelope in which the plasma temperature is of the order of the
free-fall temperature,
   \be\label{tff}
T(R) \simeq T_{\rm ff}(R)=\frac{GM_{\rm ns} m_{\rm p}}{k R}.
   \ee
Here $M_{\rm ns}$ is the mass of the neutron star, $m_{\rm p}$ is the
proton mass and $k$ is the Boltzmann constant.
The rotational energy loss by the neutron star is convected up
through the envelope by the turbulent motions and lost through its
outer boundary. The structure of the envelope and the spindown rate
of the neutron star depend on the  value of the ratio:
   \bdm
\kappa=\frac{\omega\r}{V_{\rm s}(\r)},
   \edm
where $\r$ is the magnetospheric radius, $\omega=2\pi/P_{\rm s}$
is the neutron star angular velocity and $V_{\rm s}(\r)$ is the
sound speed at the magnetospheric radius, which according to
Eq.~(\ref{tff}) is of the order of the free-fall velocity, $V_{\rm
ff}$:
    \be\label{vs}
V_{\rm s}(\r) \simeq V_{\rm ff}(\r)=\sqrt{\frac{2GM_{\rm ns}}{\r}}.
   \ee
On this basis Davies \e (\cite{dfp79}) distinguished three
sub-states of the propeller state\footnote{According to their
classification case `a' corresponds to the pulsar-like
spindown, i.e. the ejector state.}: (b)~very rapid rotator ($\kappa
\gg 1$); (c)~supersonic propeller ($\kappa \ga 1$) and (d)~subsonic
propeller ($\kappa < 1$).

In cases `b' and `c' the magnetospheric radius of the neutron
star exceeds its corotational radius,
     \bdm
R_{\rm cor}= \left(\frac{GM_{\rm ns}}{\omega^2}\right)^{1/3}.
     \edm
This means that the neutron star in these cases is in the centrifugal
inhibition regime and hence, the effective accretion onto its surface
is impossible.

In the case `d' the magnetospheric radius is smaller than the
corotational radius. In this situation the plasma being penetrated
from the base of the envelope into the star magnetic field is able to
flow along the magnetic field lines and to accrete onto the star
surface. However the effective plasma penetration into  the
magnetosphere does not occur, if the magnetospheric boundary is
interchange stable. According to Arons \& Lea (\cite{al76}) and 
Elsner \& Lamb (\cite{el76}) the onset condition for the interchange
instability of the magnetospheric boundary reads
   \bdm
T < T_{\rm cr} \simeq 0.3\ T_{\rm ff}.
  \edm
This means that the neutron star can change its state from the
{\it subsonic propeller} to {\it accretor} only when the cooling of
the envelope plasma (due to the radiation and convective motions)
dominates the energy input to the envelope due to the propeller
action by the neutron star.

Investigating this particular situation Davies \& Pringle
(\cite{dp81}, hereafter DP81) have shown that the energy input to the
envelope dominates the energy losses until the spin period of the
star reaches the break period, $P_{\rm br}$. Assuming the following
values of the neutron star parameters: the magnetic moment
$\mu=10^{30} \mu_{30}\,{\rm G\,cm^3}$ and the mass $m= 1 (M_{\rm
ns}/M_{\sun})$, and putting the strength of the stellar wind (in
terms of the maximum possible accretion rate)  $\dot{M}_0=10^{15}
M_{15}\,{\rm g\,s^{-1}}$ they estimated the value of the break period
as 60\,s.

However, putting the same values of parameters and following the
same method of calculations I found the value of $P_{\rm br}$ to be
of the order of 450\,s, i.e. by a factor of 7.5 larger than that
previously estimated in DP81. In this letter I
present the calculations and show that this result forced us to
change some basic conclusions about the origin of the long periods
X-ray pulsars.

    \section{Break period}

According to the picture presented by Davies \& Pringle the
magnetosphere of the neutron star in the state of subsonic propeller
is surrounded by the adiabatic ($p\propto R^{-5/2}$) spherically
symmetrical plasma envelope. Until the energy input to the envelope
dominates the energy losses the temperature of the envelope plasma is
of the order of the free-fall temperature, $T\simeq T_{\rm ff}$ and,
correspondingly, the sound speed is of the order of the free-fall
velocity, $V_{\rm s} \simeq V_{\rm ff}$. Under this condition the
height of the homogeneous atmosphere through out the envelope is
comparable to $R$ and thus, the envelope is extended from the
magnetospheric radius,
    \be\label{rm}
\r \simeq r_{\rm m} \equiv \left(\frac{\mu^{2}}{\dot{M}_0
\sqrt{2GM_{\rm ns}}}\right)^{2/7} =
    \ee
    \bdm
\hspace{7mm}= 1.2 \times 10^9\,{\rm cm}\  \mu_{30}^{4/7}\
\dot{M}_{15}^{-2/7}\ m^{-1/7},
       \edm
up to the accretion radius of the neutron star,
    \bdm
R_{\alpha} \equiv \frac{2GM_{\rm ns}}{V_{\rm rel}^2} = 2.7 \times
10^{10}\,{\rm cm}\ m\ V_8^{-2}.
    \edm
Here, $V_8$ is the relative velocity between the neutron star
and the stellar wind plasma, $V_{\rm rel}$, expressed in units of
$10^8\,{\rm cm\,s^{-1}}$ and $\dot{M}_0$ is strength of the stellar
wind which is expressed following DP81 in terms of
the maximum possible accretion rate:
   \bdm
\dot{M}_0 = \pi R_{\alpha}^2 \rho_{\infty} V_{\rm rel},
  \edm
where $\rho_{\infty}$ is the density of the stellar wind plasma.
Within the considered picture the envelope is quasi-static, i.e.
the mass flux through the envelope is almost zero. In this situation
the physical meaning of the parameter $\dot{M}_0$ is the rate by
which the stellar wind plasma overflows the outer edge of the envelope
compressing the envelope plasma.

The interaction between the fast rotating magnetosphere and the base
of the envelope leads to the turbulization of the envelope plasma.
The velocity of the convection motions at the magnetospheric
boundary is obviously limited as
   \be\label{vt}
V_{\rm t} \la \omega \r.
   \ee
Under the condition $\r < R_{\rm cor}$ the maximum velocity of the
convective motions is smaller than $V_{\rm s}$ and hence, the Mach
number at the base of the envelope is $M_{\rm Mach}=V_{\rm t}/V_{\rm
s} < 1$.

The rate of energy loss by the neutron star and, correspondingly,
the energy input to the envelope in this case can be expressed as (see
Eqs.~3.2.1 and 3.2.2 in DP81)
   \bdm
L_{\rm d} = 2 \pi \r^2 \rho V_{\rm t}^3.
   \edm

As it has been argued by Davies \& Pringle the cooling of the
envelope plasma is determined by the combination of convective
motions and bremsstralung radiation. In order to evaluate the
energy balance between the heating and cooling processes they
introduced the convective efficiency parameter (see also Cox \& Guili
\cite{cc68}):
    \bdm
\Gamma = \frac{\rm Excess\ heat\ content\ of\ convective\ blob}{\rm
Energy\ radiated\ in\ the\ lifetime\ of\ a\ blob}.
   \edm
Under the conditions of interest this parameter can be expressed as
(for discussion see DP81, page~221)
  \be\label{gamma}
\Gamma = M_{\rm Mach}^2 \left[\frac{V_{\rm t} t_{\rm br}}{R}\right] =
\frac{V_{\rm t}^3 t_{\rm br}}{V_{\rm s}^2 R},
    \ee
where $t_{\rm br}$ is the bremsstralung cooling time:
  \be\label{tbr}
t_{\rm br} = 6.3 \times 10^4
\left[\frac{T}{10^9\,{\rm K}}\right]^{1/2}
\left[\frac{n}{10^{11}\,{\rm cm^{-3}}}\right]^{-1} {\rm s}.
  \ee
Here, $n$ is the number density of the envelope plasma which at the
base of the envelope can be evaluated as
   \be\label{n}
n(\r) = \frac{\mu^2}{4 \pi k T_{\rm ff}(\r) \r^6}.
   \ee
As it has been shown in DP81 the cooling of the envelope
during the subsonic propeller state occurs first at its inner radius.
Thus, the energy input to the envelope due to the neutron star
propeller action dominates the radiative losses if $\Gamma(\r)
\ga 1$.

Combining Eqs.~(\ref{tff}-- \ref{n}) I find the condition
$\Gamma(\r) \ga 1$ to be satisfied if the spin period of the neutron
star is $P_{\rm s} \la P_{\rm br}$, where the break period is
    \be\label{pbr}
P_{\rm br} \simeq\ 450\ \mu_{30}^{16/21}\
\dot{M}_{15}^{-5/7}\ m^{-4/21}\ {\rm s}.
      \ee

This value of the break period exceeds the value  of $P_{\rm
br}$ presented in DP81 by a factor of 7.5 (see Eq.~4.8
in DP81). Hence the natural question about the reason of this
inconsistency arises. One of the most possible reasons is that Davies
\& Pringle have mistakenly used the value of the magnetospheric
radius: $4.4\times 10^8 \mu_{30}^{4/7} \dot{M}_{15}^{-2/7}
m^{-1/7}$\,cm (see Eq.~3.2.3 of their paper) instead of the correct
value which is expressed in their paper by Eq.~(2.5), i.e. $10^9
\mu_{30}^{4/7} \dot{M}_{15}^{-2/7} m^{-1/7}$\,cm. Taking into account
that $P_{\rm br}(\r)\propto \r^{5/2}$ one finds that the correct
value of $P_{\rm br}$ should be larger than that derived in DP81 by a
factor of 7.7, i.e. very close to the value of the break period
obtained in this letter.

    \section{Discussion}

One of the main astrophysical reasons for the investigation of the
spindown of neutron stars is the existence of X-ray sources which
display pulses with long periods (in excess of 100\,s). On the basis
of their calculations Davies \& Pringle suggested that the periods of
neutron stars spinning down due to propeller mechanism can be as
long as 100\,s only if the stars are situated in the weak stellar
wind, i.e. $\dot{M}_0 < 4 \times 10^{14}\,{\rm g\,s^{-1}}$. They also
pointed out that in this case however it is difficult to account for a
substantial population of long period pulsators.

In the light of the recalculated value of the break period obtained
in this paper (Eq.~\ref{pbr}) I find that the propeller mechanism can
be responsible for the long spin period of a neutron star even if it
is situated in the essentially stronger stellar wind:
    \be
\dot{M}_0 \la 8 \times 10^{15}\ \mu_{30}^{16/15}\ m^{-4/15}\
P_{100}^{-7/5} {\rm g\,s^{-1}},
   \ee
where $P_{100}$ is the observed spin period of the neutron star
expressed in units of 100\,s. The corresponding spindown time-scale
of the neutron star in the state of subsonic propeller is
   \be
\tau_{\rm d} \simeq 10^5\ \mu_{30}^{-2}\ m\ I_{45}\ P_{100}\  {\rm
yr},
   \ee
i.e. smaller that the characteristic evolutionary time-scale of the
early spectral type supergiants. Here $I_{45}$ is the moment of
inertia of the neutron star expressed in units of $10^{45}\,{\rm
g\,cm^2}$.

           \section{Conclusion}

The value of the break period at which the spinning down neutron star
changes its state from the {\it subsonic propeller} to {\it accretor}
obtained by Davies \& Pringle (\cite{dp81}) is underestimated by a
factor of 7.5. The incorporation of the re-estimated value of the
break period into the spindown scenario suggested by Davies \&
Pringle shows that the propeller mechanism can be responsible for the
origin of the long period X-ray pulsators even if the strength of the
stellar wind, in which a neutron star is situated, is in excess
of $10^{15}\,{\rm g\,s^{-1}}$. The analysis of the spin evolution of
a neutron star situated in the strong stellar wind will be presented
in a forthcoming paper.

 \begin{acknowledgements}
I acknowledge the support of the Follow-up  program of
the Alexander von Humboldt Foundation. The work was partly supported
by the Federal program
``INTEGRATION'' under the grant KO\,232.
\end{acknowledgements}

\end{document}